\documentclass[final, times, numbers]{elsarticle}
\setcitestyle{square, numbers}
\usepackage[colorlinks=true, citecolor=blue, linkcolor=blue, urlcolor=blue]{hyperref}
\usepackage[all]{hypcap} 
\usepackage[a4paper, total={7in, 9in}]{geometry}
\usepackage{amssymb}
\usepackage{lipsum}
\usepackage{graphicx}
\usepackage{subfigure}
\usepackage{amsmath}
\usepackage{xcolor}
\usepackage{setspace}
\usepackage{tabularx} 
\usepackage{nomencl}
\usepackage{multicol}
\usepackage{framed}
\usepackage{float}
\usepackage{booktabs}
\usepackage{etoolbox}
\usepackage{enumitem} 
\usepackage{wrapfig}
\usepackage{listings}
\usepackage{color}

\definecolor{dkgreen}{rgb}{0,0.6,0}
\definecolor{gray}{rgb}{0.5,0.5,0.5}
\definecolor{mauve}{rgb}{0.58,0,0.82}

\journal{arXiv}
\begin{document}

\begin{frontmatter}

\title{Global Estimation of Building-Integrated Facade and Rooftop Photovoltaic Potential by Integrating 3D Building Footprint and Spatio-Temporal Datasets}

\author[1]{\mbox{Qing Yu}\corref{cor1}}
\author[2]{\mbox{Kechuan Dong}\corref{cor1}}
\author[3,5]{\mbox{Zhiling Guo}\corref{cor2}}
\author[1]{\mbox{Jiaxing Li}}
\author[3,5]{\mbox{Hongjun Tan}}
\author[2]{\mbox{Yanxiu Jin}}
\author[1]{\mbox{Jian Yuan}}
\author[1,5]{\mbox{Haoran Zhang}\corref{cor2}}
\author[3,5]{\mbox{Junwei Liu}}
\author[6]{\mbox{Qi Chen}}
\author[3,5]{\mbox{Jinyue Yan}\corref{cor2}}
  
\cortext[cor1]{Equal contribution}
\cortext[cor2]{Corresponding author}

\affiliation[1]{organization={School of Urban Planning and Design, Peking University Shenzhen Graduate School},
            city={Shenzhen},
            postcode={518055}, 
            state={Guangdong},
            country={China}}

\affiliation[2]{organization={Center for Spatial Information Science, University of Tokyo},
            city={Kashiwa},
            postcode={277-8568}, 
            country={Japan}}

\affiliation[3]{organization={Department of Building Environment and Energy Engineering, The Hong Kong Polytechnic University},
            city={Hong Kong},
            country={China}}

\affiliation[5]{organization={International Centre of Urban Energy Nexus, The Hong Kong Polytechnic University},
            city={Hong Kong},
            country={China}}

\affiliation[6]{organization={School of Geography and Information Engineering, China University of Geosciences (Wuhan)},
            city={Wuan},
            country={China}}
        
\begin{abstract}
This research tackles the challenges of estimating Building-Integrated Photovoltaics (BIPV) potential across various temporal and spatial scales, accounting for different geographical climates and urban morphology. We introduce a holistic methodology for evaluating BIPV potential, integrating 3D building footprint models with diverse meteorological data sources to account for dynamic shadow effects. The approach enables the assessment of PV potential on facades and rooftops at different levels—individual buildings, urban blocks, and cities globally. Through an analysis of 120 typical cities, we highlight the importance of 3D building forms, cityscape morphology, and geographic positioning in measuring BIPV potential at various levels. In particular, our simulation study reveals that among cities with optimal facade PV performance, the average ratio of facade PV potential to rooftop PV potential is approximately 68.2\%. Additionally, approximately 17.5\% of the analyzed samples demonstrate even higher facade PV potentials compared to rooftop installations. This finding underscores the strategic value of incorporating facade PV applications into urban sustainable energy systems.

\end{abstract}

\begin{keyword}
\def\sep{; }
Urban energy system\sep
Renewable energy\sep
Solar energy\sep
Building-integrated photovoltaic potential\sep
3D building footprint\sep
Geo-information science 

\end{keyword}


\end{frontmatter}

\section{Introduction}

In response to pressing climate challenges, the Paris Agreement delineates the critical objective of limiting global warming to below 2 degrees Celsius above preindustrial levels, primarily through the implementation of Nationally Determined Contributions (NDCs), which are crucial in mobilizing innovative energy solutions across nations to achieve this ambitious goal \cite{ParisAgreement2015}. 
According to the International Energy Agency (IEA), 57 countries are committed to decarbonization strategies, and 179 countries have set ambitious goals for the expansion of renewable energy at various governmental levels \cite{2}.
In pursuit of these objectives, distributed renewable energy systems are gaining more attention \cite{ accelerating,yan2019city}, particularly reflected in the increasing deployment and installation of urban PV systems \cite{PVdata,BIPVDigitalization,SolarEnergyPlanning}.
Beyond mere quantitative increases \cite{PVCost,integrated}, urban PV systems are evolving from conventional Building-Attached PV (BAPV) towards more integrated and aesthetically appealing BIPV solutions \cite{PVWindow,PVWindow2,bipvcalculation}.
The market value of BIPV was USD 19.82 billion in 2022 and is projected to reach USD 89.8 billion by 2030, a 453\% increase, reflecting its potential to alleviate energy shortages and gain increasing recognition \cite{natureBIPV2017, GVR}.


Existing global evaluations of the PV potential have unveiled the substantial capacity of building rooftop PV to address energy challenges \cite{Roof,roofpotential,suri2020global}. 
An often underestimated fact is that in densely populated large cities with a high demand for energy supply and numerous high-rise buildings, the surface area on the urban facades is often much larger than that on the roofs \cite{36}.
As a supplementary option to rooftop solar panels, facade-integrated solar panel technology can be a potentially plentiful resource to address energy supply in urban areas facing energy shortages \cite{3, Couderc2018}.
In this context, it is crucial to evaluate the comprehensive PV potential of both facades and rooftops. 
It not only guides the next generation of urban design but also informs future decisions about urban carbon neutrality pathways, offering a strategic blueprint for sustainable development in energy-demanding urban environments \cite{wheeler2022photovoltaic,chen2022improved,CV}.

The evaluation of the PV potential in urban areas poses notable challenges due to the varied and intricate architectural context, further complicated by factors such as time, location, solar exposure, and shading effects induced by urban morphology \cite{5,ereview}.
The commonly used method to estimate the potential of PV energy involves the use of remote sensing technologies and GIS data \cite{ZHU2023100129} to identify suitable areas on roofs of buildings where PV modules can be installed and to analyze the solar radiation received by these modules \cite{8,CVSEG}. 
Advanced studies further expand and detail their methods by introducing architectural characteristics categorized\cite{13,SolarEnergyPotential}, skyline analysis \cite{NE}, and Light Detection and Ranging (LiDAR) technology for 3D architectural analysis\cite{LiDAR,gagnon2016rooftop,16} to allow precise calculations of solar radiation and detailed estimation of the PV yield.
Recently, software tools such as the System Advisor Model (SAM) \cite{sam_nrel}, PVGIS \cite{PVGIS}, and SolarGIS \cite{solargis2024} have been developed to predict PV system output by simulating regional solar irradiance and weather conditions \cite{sam_nrel, 10}.
However, the aforementioned methods face limitations in effectively analyzing shading effects resulting from urban 3D architectural structures. These limitations hinder the ability to conduct a thorough and comprehensive evaluation of the PV potential of building facades and rooftops, as well as addressing broader issues at the city or global level.

Recent progress in GIS applications has gradually reduced the difficulty of acquiring large-scale detailed building footprint data for urban areas \cite{naturelidar1, prumers2022lidar}. 
Utilizing high-resolution 3D building footprint data from open-source platforms like OpenStreetMap \cite{25} in combination with comprehensive worldwide meteorological data from databases such as the National Solar Radiation Database (NSRDB) \cite{34} provides a feasible and valuable opportunity to comprehensively evaluate the BIPV potential of buildings within urban 3D architectural contexts and expand such analysis to a global scale.

To address the critical need for evaluating the BIPV potential within global cities, this study introduces a comprehensive solution that integrates multi-source 3D urban building footprint data for a global assessment of BIPV potential. 
Utilizing advanced shadow simulation techniques that capture the distinctive shadowing dynamics of urban settings, the provided solution is capable of estimating facade and rooftop PV potential from individual buildings up to cities and the global level.
Based on the proposed solution, this study investigated and explored the influence of urban 3D morphology on the potential for BIPV and conducted a global analysis in 120 cities worldwide, showcasing facade PV's substantial energy potential. 
In this study, an open source toolkit called \textit{pybdshadow} is developed and proposed that offers the capability to produce PV potential estimation and output results into widely used geographic data formats such as Shapefile and GeoJSON, facilitating easy integration into existing geospatial analysis workflows.
Our comprehensive solution enables a detailed evaluation of the PV potential within the 3D architectural context of urban areas.
This could play a crucial role in advancing the transition of cities toward achieving carbon neutrality goals and fostering sustainable development in energy-intensive urban environments.

\begin{figure*}
\centering 
\includegraphics[width=1\textwidth]{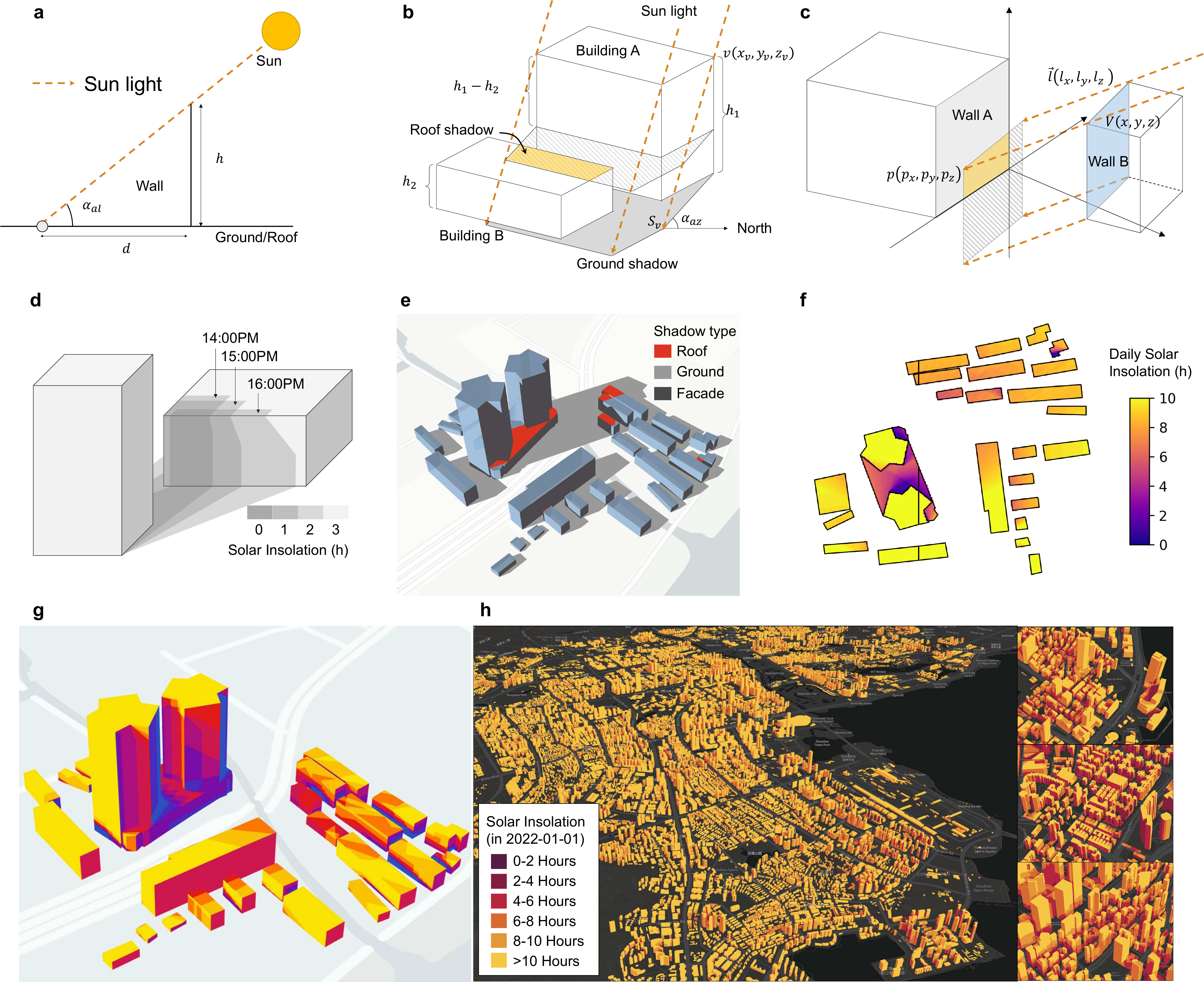}
\caption{\textbf{Computation of solar insolation within 3D architectural contexts.} \textbf{a} Shows the geometry relationship of a wall with its projection on the ground in a sun position. \textbf{b} illustrating the geometric relationship in which Building A casts a shadow over the roof of Building B. \textbf{c} depicts how Wall B casts a shadow onto Wall A. \textbf{d} shows how to represent and estimate the detailed solar insolation time on the building roof and facade in a period by overlaying the building shadows, this will generate a number of smaller polygons on each building. \textbf{e} gives an example showing the estimation results of ground, facade and roof shadows for a city block at a given time. \textbf{f} and \textbf{g} visualize the daily solar insolation time of this city block on January 1, 2022, in both the roof and 3D views. \textbf{h} shows an example of solar insolation time for 23,904 buildings in Shenzhen, China, as of January 1, 2022.} 
\label{figure1}%
\end{figure*}

\section{Solar insolation estimation based on shadow casting}

In a 3D urban setting with complex architectural forms, a crucial aspect of assessing the potential of BIPV is the examination of building shadows, which play a significant role in determining the access to solar insolation. To address this, we propose a shadow simulation technique based on 3D urban building footprint data, which enables the modeling of urban shadows at any given geographical location at any given time with geometry projection calculation. 

The substantial component of the model includes creating a detailed shadow profile for each building at a given time and location, focusing on analyzing shadows cast on rooftops and facades generated by buildings.
Initially, the process involves meticulously breaking down the structure into individual facades and establishing a plane equation for each targeted facade in a 3D space. 
By considering the sun's altitude \(\alpha_{al}\) and azimuth \(\alpha_{az}\) (which are calculated using Suncalc\cite{SunCalc}, see Fig.\ref{figure1}a), the computation of shadow projections includes integrating vertex coordinates to accurately delineate the geometric shapes of the shadows cast by each facade individually. A pivotal aspect of this shadow simulation is determining the projection point \((S_v)\) on the ground for any given point \((x_v,y_v)\) on the wall, considering the sun's altitude and azimuth angles. This is expressed as:
\begin{equation}
S_v = (x_v + h\cot(\alpha_{al})\cos(\alpha_{az}), y_v + h\cot(\alpha_{al})\sin(\alpha_{az}), 0)
\end{equation}
where \(h\) is the distance from the wall point to its projection on the ground or rooftops (Fig.\ref{figure1}b).
Fig.\ref{figure1}e illustrates the shadow conditions estimated in an urban block at a specific moment (top view).

Subsequently, for shadows on the facades (Fig.\ref{figure1}c) of the buildings, the calculation \ref{facade1} and \ref{facade2} is employed. The plane equation of wall \(X\) is denoted by \(Ax + By + Cz + D = 0\). Here, \(V = (x, y, z)\) represents the coordinates of a point on a potential shadow-casting wall, while \(P = (P_x, P_y, P_z)\) denotes the coordinates of the projected shadow point on the target wall \(X\). The solar light vector is given by \(\vec{l} = (l_x, l_y, l_z)\), which indicates the direction of sunlight. The projection calculation is defined by:

\begin{equation}
t = -\frac{A x + B y + C z + D}{A l_x + B l_y + C l_z}
\label{facade1}
\end{equation}

And the projected point \(P\) is calculated as:

\begin{equation}
P = (x + t l_x, y + t l_y, z + t l_z)
\label{facade2}
\end{equation}

From shadow casting to solar insolation on building surfaces, a comprehensive analysis of shadow dynamics over time is required. 
This involves establishing temporal boundaries during the day, dividing them into time intervals ($\delta_t$), and monitoring the movement and overlap of shadows at various time points to gather solar insolation data.
To avoid excessively long shadows caused by a low solar altitude angle $\alpha_{al}$, a buffer time interval is set to limit the calculation of temporal boundaries at sunrise and sunset.
Subsequently, shadows at different timestamps are determined using 3D geometric overlay, partitioning a wall or roof into multiple small surface areas with unique solar radiation properties, as depicted in Fig.\ref{figure1}d. 
By adjusting the time increments and buffer time frame,
one can manage the precision and computational complexity of the model. 
Fig.\ref{figure1}f displays the solar radiation pattern throughout a day with a 15-minute time interval and a 30-minute buffer interval in a top view, while Fig.\ref{figure1}g presents the results at 1-hour intervals in a 3D representation. Despite the differing resolutions in the two figures, both illustrate the same overall pattern.
With the setup of 1-hour intervals, this approach can be quickly applied to conduct broader assessments at the city level. For example, in Fig.\ref{figure1}h, the solar irradiation conditions in Shenzhen on January 1, 2022 are illustrated.
More details on the derivation of spatial relationships of shadow computation models are provided in the Methods section. 
Validation of the shadow casting estimation are provided in supplementary information.

\section{PV power generation estimation model}

The model then refines solar radiation information by weighting solar insolation at each time point based on the solar irradiance intensity on building surfaces. 
For each time interval, the calculation of the total intensity of solar radiation incorporates the diversity of solar irradiation conditions in various segments of the wall. This includes the effects of direct solar irradiation, sky diffuse, and ground-reflected components \cite{NE}. Irradiance determination involves the amalgamation of Direct Normal Irradiance (DNI), Diffuse Horizontal Irradiance (DHI), and Global Horizontal Irradiance (GHI), supplemented by shadow simulation techniques, which serve to determine the solar irradiance of different planes.

The total irradiance received by the Plane of Array (POA) during each time interval is determined by summing the contributions from different sources of irradiance \cite{POA}:

\begin{equation}
POA= G_{Dir} + G_{Dif} + G_{Ref}
\label{POA}
\end{equation}
where \(G_{Dir}\) represents the direct light irradiance, \(G_{Dif}\) represents the diffuse light irradiance, and \(G_{Ref}\) represents the reflected light irradiance. This method integrates these factors to accurately determine the solar irradiance on various surfaces, effectively simulating real-world solar conditions.

Our evaluation of PV system performance in 3D architectural settings calculates surface solar irradiance and integrates
variables: the tilt angle of PV modules mounted on building rooftops \cite{30}, ambient temperature \cite{31}, atmospheric conditions \cite{32}, and the energy conversion efficiency \cite{33}. Utilizing the pvlib software model \cite{pvlib}, our approach simulates realistic PV outputs from POA irradiance under various sunlight conditions, incorporating these critical factors. Furthermore, given the constraints associated with installation, this study proposes adjusting the conversion efficiency of facade-based PV systems to 68\% of the efficiency observed in conventional rooftop PV systems \cite{BIPVEFF}.

In addition to these considerations, the model employs shadow simulation at different time intervals to analyze the PV outputs for various building orientations and configurations. This dynamic shadowing effect is crucial for predicting realistic energy generation patterns throughout the day. By accounting for the temporal variations in shadowing, the model offers a more comprehensive understanding of the potential energy yields from PV systems under typical operational conditions. This enhanced simulation capability allows for the optimization of PV system placement and configuration to maximize energy efficiency and sustainability in urban architectural designs.
Further elaboration on the derivation of the PV power generation estimation model is provided in the Methods section.






\section{Global-scale building integrated facade and rooftop PV potential estimation}

\begin{figure*}
\centering 
\includegraphics[width=1\textwidth]{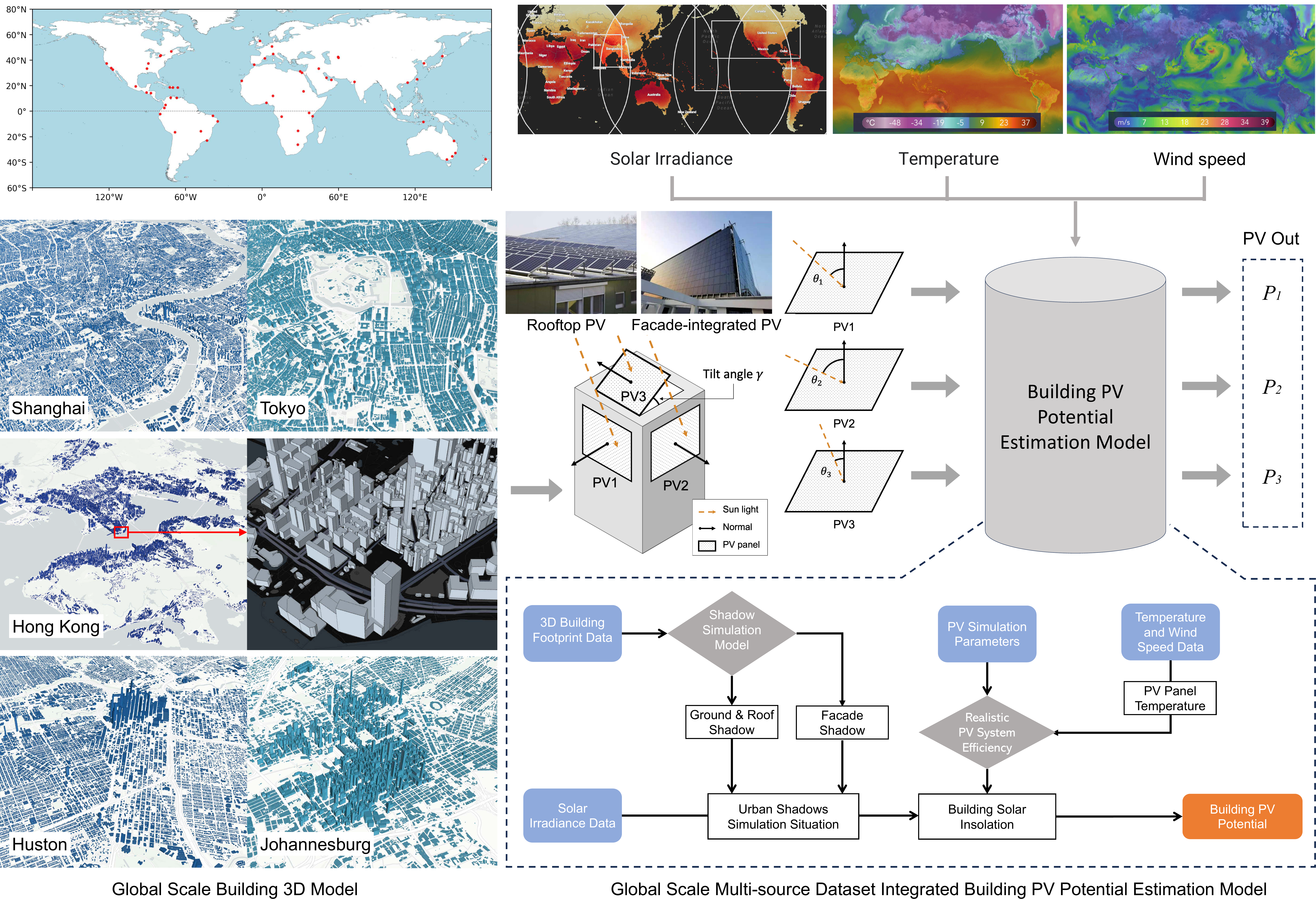}
\caption{\textbf{Global-scale building-integrated facade and rooftop PV potential estimation framework.}  By integrating building footprint data with global solar data, the model estimates rooftop and facade solar irradiance first. Then, it uses PV module parameters and weather data to estimate the PV output for each building surface. The model framework enables the estimation of PV potential at the building surface level for individual buildings and can be further extended to the estimation of PV potential of urban blocks on a global city scale.
} 
\label{figure2}%
\end{figure*}

To extend the model to a global scale, the proposed method evaluates PV potential by integrating global 3D building footprint data and meteorological data, including solar irradiance, temperature, and wind speed.
The 3D building footprint data utilized in this research on a global scale were gathered and derived from map tiles sourced from various datasets offered by Mapbox, OpenStreetMap, Tencent Map, and other map providers. 
Related studies have compared the quality assessment of these datasets and have shown that although some architectural details are missing, overall, they exhibit a high level of completeness and semantic accuracy compared to real architectural information \cite{fan2014quality}. 
The building data source underwent a series of data-cleaning operations to ensure its usability, including standardizing data formats, removing outliers, merging polygons, and integrating it into a comprehensive dataset. 
Solar radiation and weather data used in this study on a global scale are collected from the NSRDB. Depending on the geographical location, the spatial resolution of the data utilized is 2 to 4 km, and the temporal resolution is 5 to 15 minutes \cite{habte2017evaluation,xie2016fast}.

The overall technical framework is illustrated in Fig.\ref{figure2}. 
Using the model introduced previously, building footprint data are introduced to estimate the shadowing conditions of roofs and facades and then integrate with global solar irradiance data to support the estimation of building solar irradiance. 
Then, the power generation parameters of the PV modules and meteorological data are input into the PV power generation estimation model to estimate the PV yield of each roof and facade. 
Given the different types of PV modules and the installation conditions required between building facades and rooftops, different parameters are assigned to each PV module.
Finally, the model aggregates the PV potential of all surfaces, allowing for the evaluation of rooftop and facade PV potential at the individual building, urban block, and global city levels.

\section{Validation study in individual buildings}


In this section, a comprehensive validation and analysis is performed at the building level to illustrate the evaluation of the PV potential.
To validate the evaluation of PV potential, we compared the computational results with measured PV power generation data from an open-source PV generation dataset within a localized area. The dataset includes measured PV power generation data and on-site weather data collected from 60 grid-connected rooftop PV stations on the campus of The Hong Kong University of Science and Technology (HKUST) over 2021-2023, with the PV power generation data recorded at 5-minute intervals \cite{lin2024high}.
Correspondingly, we obtained the Building footprint data for this area from Tencent Maps, which encompasses 149 buildings with information on building heights. 

\begin{figure*}
\centering 
\includegraphics[width=1\textwidth]{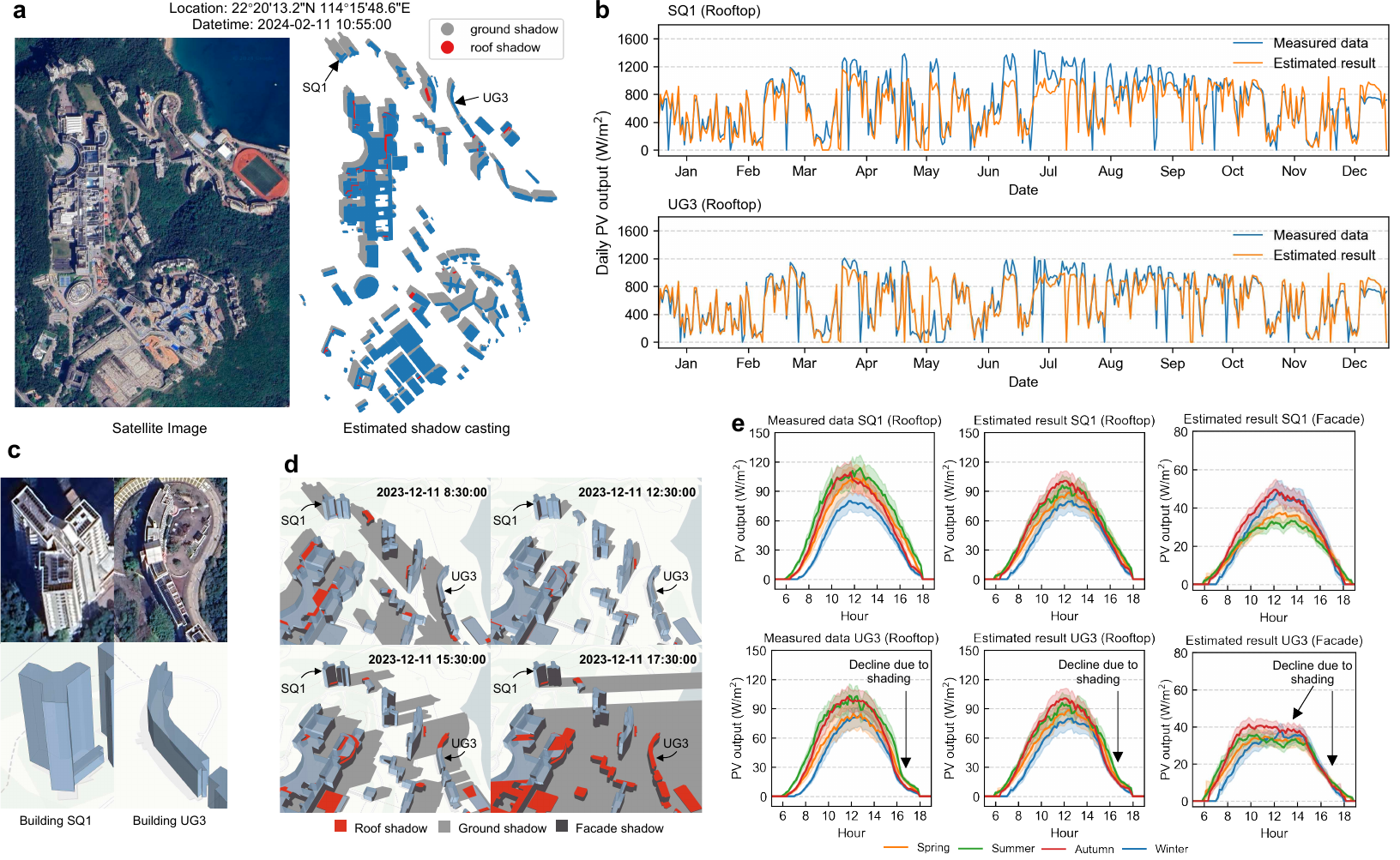}
\caption{\textbf{Validation and analysis of building-level PV potential by comparing with measured PV output data} \textbf{a} shows the comparison of the satellite image and the shadow casting estimated by building footprint at the same timestamp. \textbf{b} compares the rooftop daily measured PV output with the estimated PV potential for two selected buildings, SQ1 and UG3. \textbf{c} presents the satellite image and the building footprint data for the two buildings. \textbf{d} shows the shadow cast over the case region at different times in 2023-12-11, where the roof of building SQ1 is not blocked by shadows during the day, while the roof of building UG3 is blocked by the shadow of the adjacent building at around 17:30 in the afternoon. \textbf{e} compare the measured rooftop PV output with the estimated rooftop PV potential of the building SQ1 and UG3, where we can clearly see that the power generation efficiency of UG3 during the afternoon hours of 16:00-18:00 is significantly lower than that of SQ1, and this trend is well reflected in our estimated result.}
\label{figure3}%
\end{figure*}

To validate the shadow-casting outcomes, we compared the estimated shadows with those visible in the satellite images sourced from Google Maps.
By inputting the building footprint data and the timestamps of the satellite images into our model, we estimated the shadows cast on the ground, rooftops, and facades of the buildings, as illustrated in Fig.\ref{figure3}a.
It should be noted that the building footprint data represents a simplified version of the actual building shapes and heights, which may lead to some discrepancies.
Despite these potential data inaccuracies, the comparison reveals that the shadow estimations closely align with the satellite images, demonstrating that the model can accurately estimate real-world conditions.

The PV potential estimation was carried out daily throughout the year 2022, where the meteorological datasets were also provided from this dataset. 
For comparison, both the measured data and estimated PV potential are aggregated to the same time interval (10 minutes) and converted to the unit of W/m$^2$ for efficiency comparison of PV generation. 
Focusing on two specific buildings for detailed discussion, Building SQ1 is a high-rise structure while Building UG3 is relatively lower, the satellite image and building footprint data of these two are shown in Fig.\ref{figure3}c. 
Fig. \ref{figure3}b compares the estimated result with the measured daily PV output of the two buildings, which shows a similar trend throughout the year.
The result shows the fluctuations in maximum daily generation efficiency, with considerable variability between different dates within the year.
This variability is primarily due to the consideration of real-time weather conditions, including solar irradiance, temperature, and wind speed, all of which change throughout the day, thereby influencing the efficiency of PV systems. 
Fig.\ref{figure3}e illustrates the average PV generation potential for the facades and rooftops of these buildings across different seasons, presenting different shading conditions. 
For Building UG3, the rooftop is significantly obstructed by the taller surrounding structures during the twilight hours in most of the day in the year, as shown in Fig.\ref{figure3}d. 
This is reflected in Fig.\ref{figure3}e, where the rooftop PV potential during the late afternoon hours is notably lower for Building UG3 compared to Building SQ1, and this trend is well reflected in our estimated result.
By comparing the model results, each square meter generates 189.423 kWh annually (without shading) and 184.657 kWh annually (with shading), demonstrating that shading on UG3 reduces the annual PV output by approximately 2.51\%. 
This scale is not large because the shading mainly occurs during twilight time, where the PV output is relatively low in the day.

As for the intraday and annual trends in PV potential, it is evident that, due to Hong Kong's location in the Northern Hemisphere and north of the Tropic of Cancer, solar irradiation is generally higher in summer than in winter. 
Such seasonal variations are crucial for planning and designing BIPV systems, as they assist in optimizing system configurations to meet energy demands across different seasons. Beyond merely observing the magnitude of PV potential differences between summer and winter, the temporal dynamics of PV potential also vary with the changing seasons. 

In addition to weather-related factors, the shape of buildings and the layout of building clusters profoundly influence the ultimate PV potential. Modeling to understand the precise variations in PV potential not only facilitates the planning and installation of PV systems but also enhances resource allocation and optimization. This modeling approach enables more intuitive and clear analyses of the performance in terms of electricity generation quantity and stability of urban architectural PV systems, providing effective guidance for targeted research and design in subsequent studies.

\section{PV potential in urban blocks}

\begin{figure*}
\centering 
\includegraphics[width=0.8\textwidth]{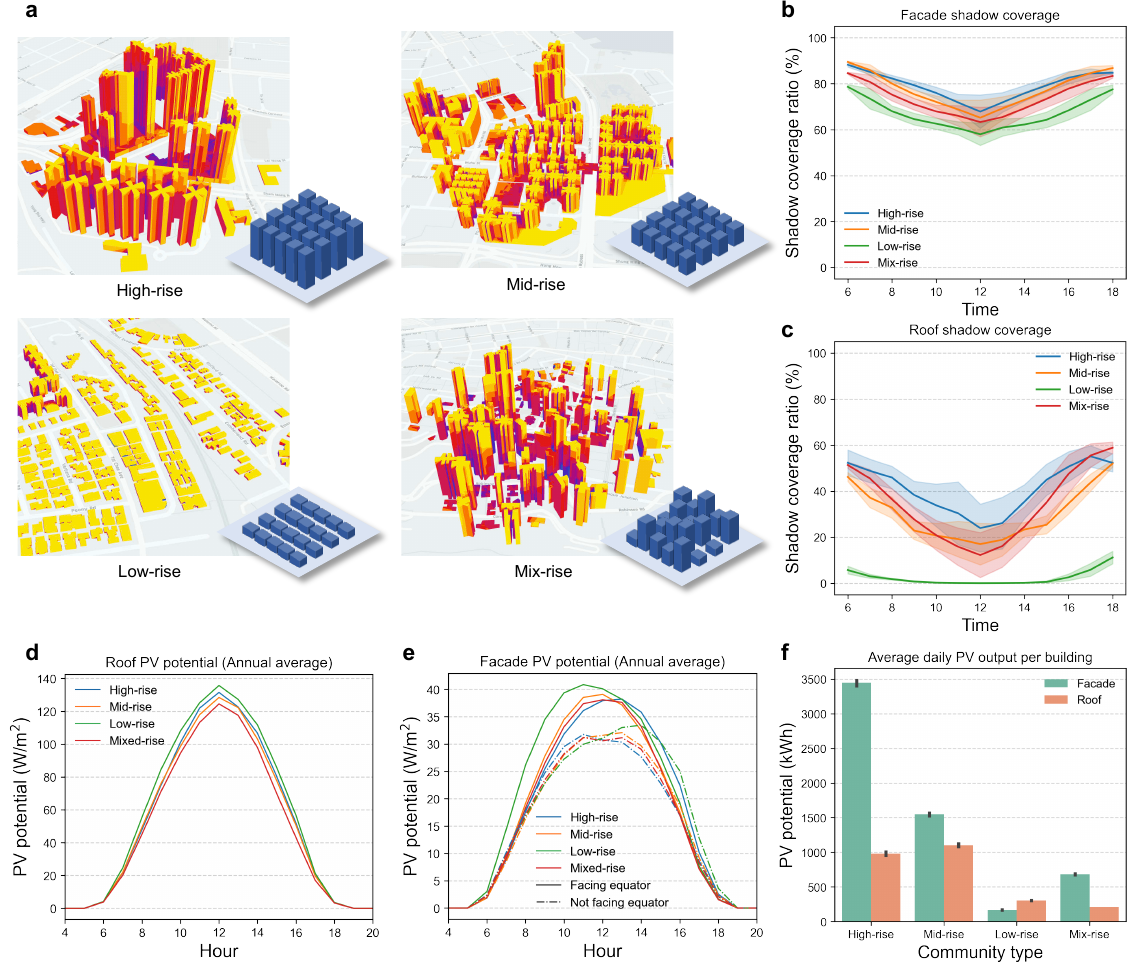}
\caption{\textbf{BIPV potential of four types of 3D urban spatial morphology in Hong Kong} \textbf{a} Illustrates the four categories of city blocks, each characterized by a unique 3D spatial configuration, accompanied by an estimation of their respective PV potential on building roofs and facades. \textbf{b} and \textbf{c} provide detailed insights into the shadow coverage ratio affecting both the facades and rooftops across these diverse city blocks. 
\textbf{d} and \textbf{e} show the estimated roof and facade-integrated PV potential, with shadow casting between the buildings considered. In \textbf{e}, the solid lines represent facades oriented towards the equator, while the dashed lines indicate facades not oriented towards the equator.
\textbf{f} provides a comparative analysis of the daily average PV power generation per building across the four city blocks, assuming that all available facade and rooftop surfaces are equipped with PV panels.}
\label{fig4}%
\end{figure*}

Urban architectural contexts in different areas may exhibit distinctly diverse 3D spatial configurations even within the same metropolis, and these variations significantly impact the potential of BIPV for energy generation. 
Another experiment in this study selected 4 city blocks, each measuring 500 m $\times$ 500 m, within the same city (Hong Kong) to investigate such variations.
These blocks encompass multiple buildings of different types, culminating in a variety of architectural 3D spaces\cite{stewart2012local}: high-rise, mid-rise, low-rise, and mixed-type structures (see Fig.\ref{fig4}a).
The calculations were carried out daily throughout the year 2020, with a time resolution of one-hour intervals.

Fig.\ref{fig4}b and Fig.\ref{fig4}c illustrate that compared to low-rise and mid-rise buildings, high-rise and mid-rise buildings have significantly higher shadowing ratios. 
At any given moment, a substantial portion of all building types' facades is in shadow, with the shadow ratio consistently exceeding that of roof shadows. 
Consequently, the overall efficiency of power generation from facades is notably lower than that from roofs. 

In the analysis presented in Fig.\ref{fig4}d and Fig.\ref{fig4}e, conducted during the peak sunlight hours from 9 AM to 3 PM, it was observed that low-rise buildings achieve the highest daily PV potential of the rooftop, reaching 927.95 $W/m^2$, with a shadow coverage of only 2.07\% during daytime. 
Conversely, high-rise buildings exhibit the lowest daily PV potential, measured at 771.76 $W/m^2$, with a shadow coverage of 39.61\% during daytime. 
This results in a PV potential that is only 83.16\% of that observed in low-rise buildings, a disparity primarily due to the increased effect of shadowing.
Specifically, the shadow coverage increased from 2.07\% to 39.61\%, which corresponds to a 16.84\% reduction in average daily PV potential. 

In the four area scenarios selected for this study, changes in urban building morphology exert a relatively minor influence on the overall PV potential of facades when compared to rooftops. The most significant daily variation was a 7.69\% decrease in facade PV potential, corresponding to a 7.94\% increase in shadow coverage, underscoring the more pronounced effect of rooftop configurations on PV performance relative to facades. 
One significant factor is that, at any given time, approximately half of the building facades are not exposed to direct sunlight, which limits the impact of urban morphology changes on specific sections of the facades. In this study, we observed that variations in PV potential for facades facing the equator under different urban morphological scenarios could reach up to 8.98\%, whereas facades not oriented towards the equator showed a comparatively lower variation of 6.31\%. Furthermore, within the same urban configuration, the increase in PV potential for equator-facing facades can be as high as 17.35\%, with an average increase of 15.17\%, compared to facades that do not face the equator. This indicates that the orientation of building facades, in conjunction with variations in urban morphology, exerts a profound influence on optimizing the PV potential of facades.

Another essential factor is the geometrical differences in sunlight exposure between rooftops and building facades in Hong Kong. Building facades typically have a smaller incidence angle with sunlight than rooftops. Particularly around noon, when solar radiation is most intense, the angle between building facades and direct sunlight can be very small, potentially approaching 0 degrees. As a result, the direct light irradiance on facades is significantly lower than on rooftops during peak solar radiation times. Consequently, variations in the shadowed areas on facades have a lesser impact on their PV potential. Furthermore, the conversion efficiency of PV systems installed on building facades is generally lower than those on rooftops. This implies that, even under identical solar radiation conditions, the PV potential of facades is less than that of rooftops. 

In urban environments characterized by high-rise, mid-rise, and mixed-rise buildings, facades can present even more tremendous PV potential than rooftops due to the larger surface areas available for installing PV systems. This advantage, however, is often constrained by the building's structural orientation and spatial positioning, which can limit the effectiveness of facade PV installations. Therefore, by analyzing and optimizing urban building morphology and facade PV layouts, it is possible to unlock even more significant PV potential from building facades, which is crucial in enhancing urban energy production.

\section{PV potential in global cities}

\begin{figure*}
\centering 
\includegraphics[width=1\textwidth]{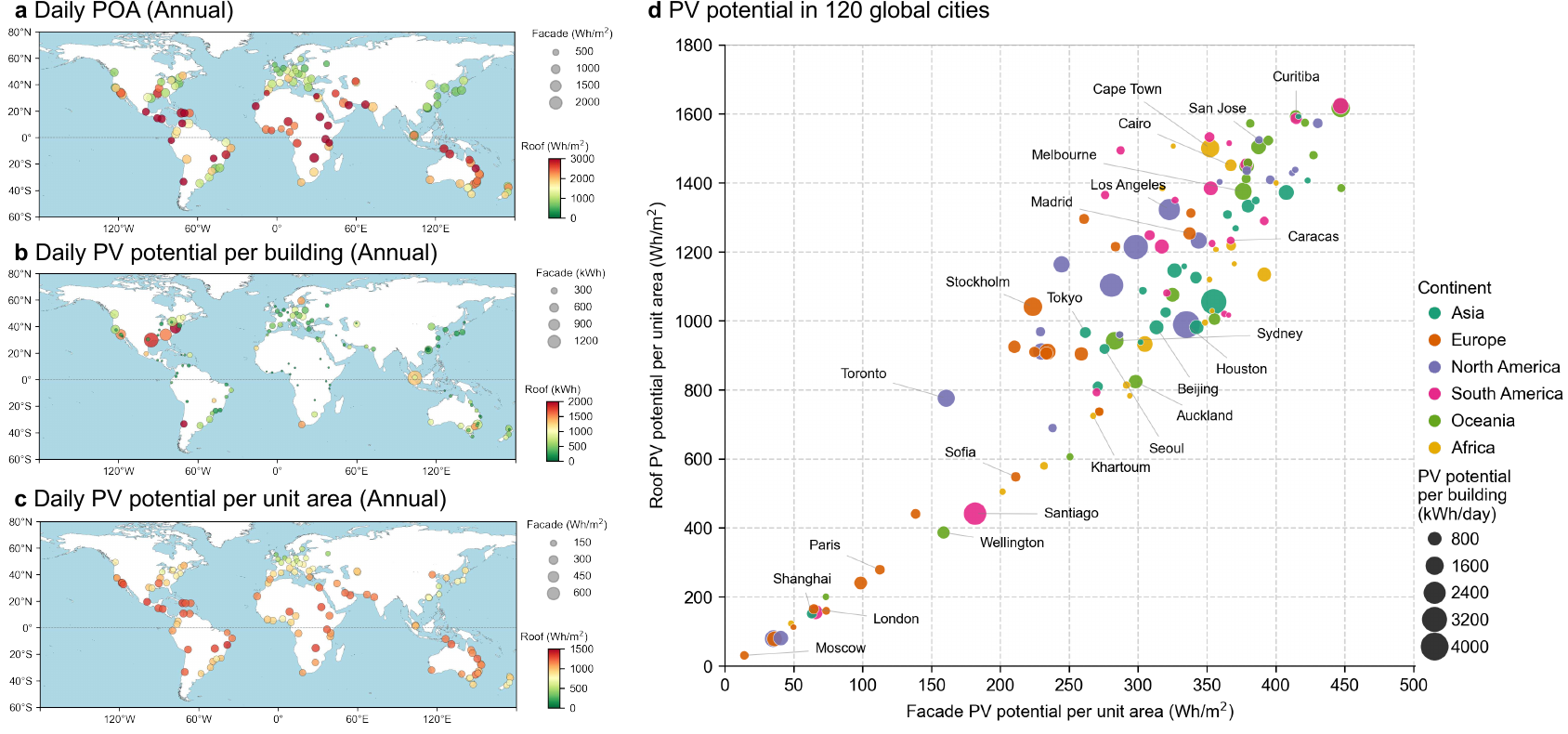}
\caption{\textbf{Comparative analysis of BIPV in central urban areas of 120 global cities.} 
The proposed methodology has been implemented across urban centers in 120 typical cities worldwide, each featuring a 1 km $\times$ 1 km area for the assessment of BIPV potential. \textbf{a} shows the estimated Plane of Array Irradiance (POA) of both roof and facade in the 120 cities, each figure uses color to represent rooftop PV and marker size to represent facade-integrated PV. \textbf{b} depicts the average daily PV potential in kWh per building, while \textbf{c} displays the daily PV potential per unit of area. \textbf{d} presents a scatter plot illustrating the PV potential of 120 cities across various continents.
}
\label{fig5}%
\end{figure*}

This research involved the selection of 120 cities (20 cities on each continent) worldwide to assess the PV potential within a 1 km $\times$ 1 km core urban zone of each city on the first day of every month throughout 2019, with a time resolution of one-hour intervals. The result is shown in Fig.\ref{fig5}.
The simulation for adjusting the tilt of the rooftop PV panels was based on the latitude of the specific city where they are situated, ensuring that all panels are uniformly facing the equator for optimal performance.
Detailed information regarding the selection criteria and list of cities has been systematically compiled in supplementary information. 

Latitude plays a pivotal role. Within the scope of tested nations, 70\% of the countries showcasing the 20 highest rooftop PV potential per unit of area fall within the 10-20 degree latitudinal band. Furthermore, around 73.08\% of cities situated within this specific latitude range are among the top 25\% in terms of their rooftop PV potential, underscoring the significant correlation between latitude and solar energy capacity.
In contrast, the geographical distribution of facade PV potential per unit area exhibits a more uniform dispersion across latitudes within the 40-degree range. Analysis of the top 20 regions reveals a nuanced latitudinal distribution: 25\% of these high-performing regions are situated within latitudes lower than 10 degrees, 35\% are located within the 10-20 degree latitudinal band, 15\% fall within the 20-30 degree range, and 20\% are positioned within the 30-40 degree latitude, indicating a relatively equitable distribution of facade PV potential across different latitudinal zones.

This variation can be attributed to the geometric orientation of the facades relative to the sun's path. While low-latitude areas receive intense solar radiation, facades, which are perpendicular to the ground, encounter a smaller angle of incidence from sunlight during peak radiation periods, thereby reducing their efficiency in capturing direct solar radiation. Conversely, as latitude increases and solar radiation diminishes, the angle at which sunlight strikes the facades becomes more acute, thereby enhancing their solar capture efficiency. Thus, within the 40-degree latitude range, facade PV potential is characterized by significant dispersion, illustrating the complexity inherent in the PV potential of facades.
This distribution underscores the substantial impact of geographical location on the efficiency of PV installations, with a clear advantage observed in latitudes closer to the equator for facade-based solar energy capture.

Moreover, a high PV potential per unit area does not necessarily translate into a high overall PV potential for buildings, as is shown in Fig.\ref{fig5}a and Fig.\ref{fig5}c.
Architectural characteristics and building forms prevalent in different cities emerge as significant determinants of the overall PV potential. 
This is exemplified by cities such as Washington D.C. in the United States, Stockholm in Europe, and Santiago in Chile, which, despite showing moderate roof-based daily conversion efficiencies, exhibit substantial rooftop PV potentials, where their daily PV potential is 2096.9, 1425.5, and 2463.8 $kWh$. 
In cities such as Houston, Singapore, and Toronto, while the PV potential per unit area of individual building facades is observed to be moderate, overall architectural facades demonstrate a substantial PV integration potential, with respective performances of 2645.6, 2617.7, and 1138.7 $kWh$. 
These data highlight the paramount importance of holistic architectural design and orientation in optimizing solar energy capture efficiency.

On balance, rooftop PV systems typically outperform facade-based systems in energy efficiency across most territories. Yet, in cities with extensive facade areas, the strategic utilization of these surfaces significantly boosts the overall PV output. Out of 120 cities surveyed, the average ratio of facade PV potential to rooftop PV potential is approximately 68.2\%. 21 cities, accounting for 17.5\% of the sample, exhibit facade PV potentials that exceed those of rooftop installations. Considering only the amount of radiation received by building facades and roofs, 35.9\% of the city's building facades receive more radiation than their roofs, indicating a significant potential for further development. Moreover, when analyzing the average ratio of solar radiation received by building facades to that received by roofs across 120 cities, it is observed that facades even outperform roofs in terms of solar radiation, reaching a ratio of 100.7\%. This indicates that, without accounting for the differences introduced by BIPV materials and design, building facades can exhibit PV potential comparable to that of roofs, and in certain regions, they may even significantly exceed it. Notably, cities such as Hongkong, Singapore, and Shenzhen achieve facade PV potentials of 216.8\%, 201.7\%, and 200.4\% relative to their rooftop systems, respectively. These findings underscore the significant, albeit often overlooked, potential of urban architectural facades in solar energy utilization. 

Comparing PV potential estimated in different continents, cities in Asia and North America have significant potential for PV applications due to rapid economic growth and urbanization, especially in high-rise buildings. 
The high energy demand and building density in these regions provide ample space for the promotion of BIPV technology. 
The focus of development is on efficiently utilizing building space based on existing buildings and improving energy efficiency. 
In South America, Oceania, and African cities, while the unit area potential is high, the unit building potential is relatively low, mainly due to building characteristics and city scale. 
In these developing countries, PV related technology plays a key role in energy transition and sustainable development. 
BIPV should be an important consideration in urban planning and architectural design, especially in new area development and urban renovation. 
The PV potential in Europe is significantly affected by regional and climatic differences. Cities like London and Paris have relatively low PV potential due to climatic limitations, while some cities in Southern and Northern Europe still have considerable potential. For the development of PV in Europe, differentiated strategies should be formulated based on regional characteristics, and efforts should be made to promote technological innovation and system efficiency improvement.

\section{Conclusion}

This study presents a methodology for estimating the PV potential of building facades and rooftops within the context of urban 3D morphology. By integrating building footprint models with multi-source meteorological data, the proposed approach simulates shadow casting on 3D buildings. The solution enables a comprehensive and robust evaluation of the energy yield of BIPV systems, spanning from individual buildings to a global scale, with high spatiotemporal resolution. To demonstrate feasibility, we calculated and analyzed the PV potential for individual buildings, four types of building blocks, and regions covering 1 square kilometer in 120 cities worldwide, providing a detailed analysis of the specific impacts of various factors on architectural PV potential. 

The case studies across 120 cities worldwide reveal that facade PV potential averages about 68.2\% compared to rooftops. Remarkably, around 17.5\% of the analyzed samples show higher facade PV potentials than rooftop installations, with some city regions experiencing over a 1.5-fold increase. These findings highlight the overlooked potential of building facades for PV installations. Despite current technological constraints limiting facade PV output compared to rooftop PV, facades hold significant potential, often surpassing rooftops in urban settings. Therefore, BIPV should be actively pursued in new construction and renovation projects. Furthermore, in less developed regions like Africa, although the POA and per unit area PV outputs are high, individual building outputs remain low. Consequently, these areas present substantial opportunities for promoting PV technology, potentially addressing energy challenges in developing nations. Moreover, unlike rooftop PV, facade PV output is less affected by latitude, with urban morphology and building orientation playing a crucial role, emphasizing the importance of these factors for urban planners, architects, and policymakers.

Our approach includes some simplifications, such as assuming flat or ideally tilted building roofs. Additionally, while we estimate solar irradiance reflection from the ground, reflections from neighboring facades are not fully explored. 
Although with certain limitations, the proposed methods enable quick estimation of detailed BIPV potential maps for any location, which can produce more detailed and practical guidance for practical implementation. 
Future research endeavors could enhance accuracy by integrating microclimate data, utilizing detailed models like point clouds and realistic 3D urban representations, and incorporating energy policies. Subsequent work should also explore the implementation potential, economic viability, and environmental benefits of BIPV systems to provide a comprehensive framework for sustainable urban development.

\section{Methods}

\subsection{Shadow computation}

The computational approach used to analyze the impact of shadows on rooftop solar insolation is grounded in trigonometric principles, facilitating a detailed examination of the interplay between sunlight and urban building geometries. 

Basically, to calculate the projection distance ($d$) of a shadow on the ground from a wall of height $h$, the following equation is utilized:

\begin{equation}
d = h  \cot(\alpha_{al})
\end{equation}
where $\alpha_{al}$ is the sun's altitude.

The projection point ($S_v$) on the ground for any given point on the wall, taking into account the sun's altitude and azimuth angles, is determined by:
\begin{equation}
S_v = (x_v + d\cos(\alpha_{az}), y_v + d\sin(\alpha_{az}), 0)
\end{equation}
where $\alpha_{az}$ is the sun azimuth. 

The assessment of a building's ability to cast shadows upon adjacent rooftops is quantified through the relative height difference (\(h_{r}=h_{1} - h_{2}\)). By substituting \(h\) with \(h_{r}\), it becomes feasible to calculate all potential shadow projections across rooftops of varying heights.

For facade shadow projections, for a given wall $X$, the plane equation $P_e$ is established using the coordinates of its vertices $w_1$, $w_2$, and $w_3$:
\begin{equation}
\begin{vmatrix}
x - x_1 & y - y_1 & z - z_1 \\
x_2 - x_1 & y_2 - y_1 & z_2 - z_1 \\
x_3 - x_1 & y_3 - y_1 & z_3 - z_1
\end{vmatrix} = 0
\end{equation}

The shadow projection involves calculating the position of a projected point $p$ from a point $V$ on a potential shadow-casting wall onto the target wall $X$. This is facilitated by the solar light vector $\vec{l}$ and the plane equation of wall $X$. The projection calculation is defined as:

\begin{equation}
t = -\frac{A  p_x + B  p_y + C  p_z + D}{A  l_x + B  l_y + C  l_z}
\end{equation}

And the projected point $p$ is given by:

\begin{equation}
p(x + t  l_x, y + t  l_y, z + t  l_z)
\end{equation}

This mathematical framework enables the precise determination of shadowed areas on the facades at any given moment. By integrating these computations, an accurate depiction of the shadow effects on building facades is achieved.

\subsection{Solar insolation estimation for building surfaces}

Quantification of solar irradiance on building surfaces over time requires a comprehensive analysis of shadow dynamics. This study adopts a methodical approach to calculate and visualize the solar exposure across different facets of a building's exterior by tracking the shadow movements and frequency of overlap at varying time intervals throughout the day. Employing a systematic overlay of shadow polygons generated at successive moments ($t_1, t_2,...,t_n$) within specified time intervals ($\delta_t$), this method facilitates the determination of the cumulative shadow effect on any given surface area, as illustrated in Fig \ref{figure1}(d).

Key parameters, such as padding value and time granularity, are meticulously set to tailor the shadow analysis to specific environmental conditions and temporal scales. The padding value is strategically chosen to mitigate the impact of excessively elongated shadows during periods of low sun angle, thereby refining the temporal bounds for shadow computation. This study utilizes a half-hour padding and a one-hour time granularity to map the shadow coverage with precision. Through this refined temporal framework, the frequency of shadow overlaps and the resultant shadowed polygons on various surfaces are calculated and aggregated. 

The duration of sunlight exposure for each distinct area is then derived by accounting for the overlap frequency, providing a granular view of the sunlight scenario on the building surfaces.

This section delineates the mathematical underpinnings utilized in calculating sunlight conditions on building surfaces, focusing on the computation of shadow overlays and their temporal dynamics as shown in Fig.\ref{figure1}d. 

The overlay of shadows at different times is computed to ascertain the cumulative impact on solar exposure. This involves calculating the shadows at each moment ($t_1, t_2,...,t_n$) within the time interval ($\delta_t$) and identifying the overlaid shadow polygons as follows:

The sunlight exposure duration for any given area is quantitatively assessed through:

\begin{equation}
t_p = n_p  \delta_t
\end{equation}
where $t_p$ represents the shadow duration for each newly formed area, calculated as the product of the number of overlaps ($n_p$) and the time granularity ($\delta_t$).

The total sunlight exposure duration ($t_s$) for a given area is subsequently determined by:

\begin{equation}
t_s = t - t_p
\end{equation}

This calculation provides a nuanced measure of sunlight condition, accounting for the varying degrees of shadow impact across different building surfaces and ground areas. For regions completely unobstructed by shadows, the sunlight exposure duration equates to the total daylight duration ($t$), offering a precise metric for assessing solar accessibility.

\subsection{PV Potential Estimation}

Estimation of BIPV potential leverages comprehensive urban 3D building footprint data to assess the interplay between shadow dynamics and solar irradiance on building surfaces. By integrating detailed shadow modeling with temporal weather conditions, this study utilizes solar irradiance data from the NSRDB to ascertain the DNI ($D_n$), DHI ($D_h$), and GHI ($G_h$) applicable to the urban context. This data, which takes into account elements such as cloud cover in the sky, can be utilized to accurately model the solar energy received by the PV system's surface. A critical step involves adjusting the DNI based on the angle $\theta$ between the sunlight vector and the building surfaces, thereby refining the direct solar radiation intensity received by the surfaces.

The solar irradiance received by a plane at any angle can be influenced by three types of light: direct, diffuse, and reflected. Therefore, the formula for calculating the POA irradiance \(I_r\) is as follows:

\begin{equation}
I_r = G_{Dir} + G_{Dif} + G_{Ref}
\label{POAcalculation}
\end{equation}
where the calculations for direct light irradiance \(G_{Dir}\), diffuse light irradiance \(G_{Dif}\), and reflected light irradiance \(G_{Ref}\) are given by the following formulas respectively:

\begin{equation}
G_{Dir} = D_n * \cos (AOI)
\end{equation}

\begin{equation}
G_{Dif} = D_h * \frac{1 + \cos (tilt)}{2}
\end{equation}

\begin{equation}
G_{Ref} = G_h * Albedo * \frac{1 - \cos (tilt)}{2}
\end{equation}

The angle of Incidence \(AOI\) quantifies the angle between incoming sunlight and the panel's normal. The \(tilt\) represents the panel's orientation relative to the ground, affecting sunlight exposure, and the \(Albedo\) measures the ground's reflectivity, influencing received solar irradiance \cite{AOI}.

Variability in solar irradiance across different regions of a wall may be attributed to obstructions that provide shade. For the entire surface area denoted as \(S\), which is subdivisible into discrete sections \(S_1, S_2, \ldots, S_m\), application of Equation \ref{POAcalculation} yields the respective irradiance values \(I_1, I_2, \ldots, I_m\). Consequently, the average solar irradiance\(I_u\) of unit area for a specified time interval \(\delta_i\) is expressed as:

\begin{equation}
I_u = \frac{\sum_{r=1}^{m} I_r  S_r}{S}
\label{solar}
\end{equation}

Taking into account that the energy transfer efficiency of the PV system is subject to variations induced by temperature fluctuations, this study incorporates the ambient temperature (\(T_i\)) and wind speed (\(W_i\)) as variables to approximate the realistic temperature conditions of the PV module. Subsequently, leveraging the real system parameters (\(\delta_i\)), as delineated in the supplementary information, the module employs the \textit{pvlib} library to compute the energy output (\(P_U\)) per unit area of the PV model. This process is elucidated in the equation referred to as Equation~\ref{eq:pv_output}.

\begin{equation}
P_u = \sum_{i = 1}^{n} f(I_u, T_i, W_i, \theta_{PV})
\label{eq:pv_output}
\end{equation}

Through the aggregation of \(P_u\) across various time intervals and geographical expanses, the model is capable of simulating the authentic PV potential for specified regions and periods.

\section{Data and Code Availability}
Utilizing the method introduced in this research, we created a Python package called \textit{pybdshadow}. This package includes functions for acquiring global building footprint data and for generating, analyzing, and visualizing building shadows and solar insolation maps, which are useful for estimating PV potential. The code can be found at \url{https://github.com/ni1o1/pybdshadow} and the document is available at \url{https://pybdshadow.readthedocs.io/en/latest/}.

\bibliography{cite2}
\bibliographystyle{unsrt}

\section{Acknowledgements}

This work was supported by Hong Kong Polytechnic University through projects,  P0043885 - Flexibility of Urban Energy Systems (FUES), P0047700 - International Research Centre of Urban Energy Nexus, and P0042845 - Data-driven solutions for decarbonizing transportation sector by coupling renewable energy, energy storage, and smart EV-charging, and the Japan Society for the Promotion of Science (JSPS) 21K14261 grant.

\section{Author Contributions}

Qing Yu: Writing - Review \& Editing, Conceptualization, Supervision, Methodology, Software, Data Curation, Visualization.

Kechuan Dong: Review \& Editing, Methodology, Software, Writing - Original Draft.

Zhiling Guo:  Writing - Review \& Editing, Conceptualization, Supervision, Methodology, Funding acquisition.

Jiaxing Li: Writing - Original Draft, Writing - Review \& Editing.

Hongjun Tan: Writing - Original Draft, Writing - Review \& Editing.

Yanxiu Jin: Writing - Review \& Editing, Data Curation.

Jian Yuan: Writing - Original Draft, Writing - Review \& Editing.

Haoran Zhang: Writing - Review \& Editing, Conceptualization, Supervision, Funding acquisition.

Junwei Liu: Writing - Review \& Editing, Conceptualization

Qi Chen: Methodology, Conceptualization

Jinyue Yan: Writing - Review \& Editing, Conceptualization, Supervision, Funding acquisition.

\section{Competing interests}

The authors declare no competing interests.

\end{document}